\documentclass{article}
\usepackage{amssymb}
\usepackage{amsmath,amsfonts}

\begin{document}

\begin{center}
{\Large Linear and cyclic codes over some special rings}%
\begin{equation*}
\end{equation*}%
Cristina Flaut$^{\ast }$ and Bianca Liana Bercea-Straton

\begin{equation*}
\end{equation*}
\end{center}

\textbf{Abstract. }{\small In this paper, we describe linear and cyclic
codes over the rings of the form }$R_{s,p}=\mathbb{Z}_{p}[u]/\left( f\left(
u\right) /\left( u-s\right) \right) ${\small , where }$p${\small \ is a
prime number and} $f\left( u\right) =u^{p}-u$, {\small with}  $s\in
\{0,1,...,p-1\}${\small .}%
\begin{equation*}
\end{equation*}

\textbf{Key words.} Cyclic codes, Linear codes, Hamming weight.

AMS Classifications: 94B05, 94B15.

\begin{center}
\begin{equation*}
\end{equation*}
\end{center}

\bigskip \textbf{1. Introduction}%
\begin{equation*}
\end{equation*}

The study of linear and cyclic codes defined over finite rings represents a
fundamental direction in information theory and cryptography, with crucial
applications in ensuring the integrity and security of modern data. Unlike
traditional codes constructed exclusively over finite fields (such as binary
ones), the richer algebraic structures of rings allow for a much higher
information density and superior modeling of complex communication channels.
Their theoretical importance is driven by the deep connections between
algebra and other branches of matematics ( combinatorics, algebraic
geometry) providing great tools for designing highly efficient decoding
algorithms capable of detecting and correcting multiple errors. Practically,
these codes form the foundation of current standards in data storage
technologies (such as flash memory or optical discs) and advanced
telecommunication systems, where energy efficiency and transmission
reliability are essential. Thus, their analysis remains crucial for
technological evolution and the security of global digital
infrastructures.\medskip 

In this paper, all rings $R$ are considered commutative and unitary
rings.\medskip 

\textbf{Definition 1. }i) ([AP; 05], [W; 99] ) Let $R$ be a commutative and
unitary ring. Over $R$ a \textit{linear code} $\mathcal{C}$ \textit{of length%
} $n$ is an $R-$submodule of $R^{n}$. If $R$ is a field, then a linear code
is an $\left[ n,k\right] $ code if the dimension of this $R-$subspace is $k$%
. For this case, we can associate a $k\times n$ matrix to an $\left[ n,k%
\right] $ linear code. The rows of this matrix form a basis in $\mathcal{C}$
and is called \textit{the generator matrix of the code} $\mathcal{C}$. \ If
this matrix is under the form $\left( I_{k}\shortmid M\right) $, where $%
I_{k} $ is $k\times k$ matrix and $M$ is a $k\times \left( n-k\right) $
parity matrix, used to determine the error-checking bits, this form is
called \textit{systematic}.

ii) ([AP; 05], Theorem 2.1) We consider the ring $R_{n}=R[x]/\left(
x^{n}-1\right) $. We associate to the vector $c=\left( c_{1},...c_{n}\right) 
$ the polynomial $c\left( x\right) =c_{1}+c_{2}x+...+c_{n}x^{n}$ in $R_{n}$.
In the ring $R_{n}$ a subset $\mathcal{C}$ of $R_{n}$ is a \textit{cyclic
code} if and only if $\mathcal{C}$ is an ideal of $R_{n}$.

iii) If $\mathcal{C}$ is a linear code of length $n$ over the ring $R$, for
a codeword $c=\left( c_{0},c_{1},...,c_{n-1}\right) \in \mathcal{C}$, 
\textit{the Hamming weight} $w_{H}\left( c\right) $ is the number of
coordinates $i$ in which $c_{i}$ is different from zero in the ring $R$. The 
\textit{minimum Hamming} weight of the code $\mathcal{C}$ is $w_{\text{min}%
}\left( \mathcal{C}\right) =$ \textit{min}$\{w_{H}\left( c\right) ,c\in 
\mathcal{C}\}$\textbf{.}

iv) ([GNW; 04], Theorem 2.8) The \textit{socle} of the $R$-module $%
M,Soc\left( M\right) ,$ is the sum of its minimal submodules. A finite ring
is called a \textit{Frobenius ring} if its socle is a principal ideal. The
direct product of fields is a Frobenius ring.

v) ([W; 99]) On $R^{n}$ we define \textit{the dot product} 
\begin{equation*}
x\cdot y=\overset{n}{\underset{i=1}{\sum }}x_{i}y_{i},
\end{equation*}%
for $x=\left( x_{1},...,x_{n}\right) ,y=\left( y_{1},...,y_{n}\right) \in
R^{n}$. For a linear code $\mathcal{C}$ over the ring $R$, we define \textit{%
the dual} of the code $\mathcal{C}$, denoted $\mathcal{C}^{\perp },$%
\begin{equation*}
\mathcal{C}^{\perp }=\{x\in R^{n},x\cdot y=0\text{, for all }y\in \mathcal{C}%
\}\text{.}
\end{equation*}

vi) If $R$ is a field, the generator matrix of the code $\mathcal{C}^{\perp
} $ is an $\left( n-k\right) \times n$ matrix, denoted $H$, and called the 
\textit{parity check matrix} for the code$\mathcal{\ C}$. If the generator
matrix of the code $\mathcal{C}$ is in the systematic form $\left(
I_{k}\shortmid M\right) $, then the parity check matrix of the dual code has
the form $\left( -M^{t}\shortmid I_{n-k}\right) $.

vii) ([H; 01], Theorem 1) On a finite commutative Frobenius ring $R$, if $I$
is an ideal of $R$, then $\left\vert I\right\vert \left\vert I^{\perp
}\right\vert =\left\vert R\right\vert $, therefore for a linear code defined
on the finite Frobenius rings, we can define the dual of this code.

viii) ([MK; 04 ], 235-236) In a ring $R$, the element $e\in R$ is called 
\textit{idempotent} if $e^{2}=e$. Two idempotents $e_{1}$ and $e_{2}$ are
called \textit{orthogonal} if and only if $e_{1}e_{2}=0$. If $\left(
e_{i}\right) $\medskip $_{i\in I}$ represents a family of indempotens of the
ring $R$, such that $e_{i}e_{j}=\delta _{ij}e$, where $\delta _{ij}$ is the
Kronecker symbol and $\underset{i\in I}{\sum }e_{i}=1$, then the ring $R$
can be write under the form: 
\begin{equation*}
R=\underset{i\in I}{\bigoplus }e_{i}R,
\end{equation*}%
called the \textit{Peirce decomposition}.\medskip 

\ 
\begin{equation*}
\end{equation*}

\textbf{2. Linear and cyclic codes over the ring} $R_{s,p}$%
\begin{equation*}
\end{equation*}

There are many papers dedicated to the study of diverse types of linear and
cyclic codes. In several papers from literature, were studied codes over
some special rings: over over the ring $R=\mathbb{F}_{q}+v\mathbb{F}%
_{q}+v^{2}\mathbb{F}_{q},v^{3}=v,$ where $q$ is a odd prime power, in 
\textbf{[MGF; 18]), }over the ring\textbf{\ }\ $R=\mathbb{F}_{2}+v\mathbb{F}%
_{2},$ with $v^{2}=v$, in [\textbf{ZWS; 10}], over the ring $R=\mathbb{F}%
_{p}+v\mathbb{F}_{p}+...+v^{p-1}\mathbb{F}_{p}$, where $p$ is a prime number
and $v^{p}=v\,$, in [\textbf{SS; 16}], etc. We can remark that the involved
polynomials are of the form $f\left( v\right) =v^{p}-v=v\left( v-1\right)
\left( v-2\right) ...\left( v-p+1\right) ,p$ a prime number, that means
polynomials having all $p$ distinct roots as all elements from $\mathbb{Z}%
_{p}$. From this reason, we consider a good idea the study of the linear and
cyclic codes over the rings of the form $R_{s,p}=\mathbb{Z}_{p}[u]/\left(
f\left( u\right) /\left( u-s\right) \right) $, where:

- $p$ is a prime number 

- the involved polynomial has only $p-2$ distinct roots as elements in $%
\mathbb{Z}_{p}$.

For this purpose, we consider  $f\left( u\right) $ of the form $f\left(
u\right) =u^{p}-u=u\left( u-1\right) ...\left( u-p+1\right) $. We choose an
arbitrary $s\in \{0,1,...,p-1\}$ and we define the ring $R_{s,p}=\mathbb{Z}%
_{p}[u]/\left( f\left( u\right) /\left( u-s\right) \right) $. We consider
the set $I_{s}=\{0,1,2,...,p-1\}-\{s\}$. The ring $R_{s,p}$, defined above,
has cardinal $p^{p-1}$.We denote 
\begin{equation*}
Q_{sr}\left( u\right) =\frac{f\left( u\right) }{\left( u-s\right) \left(
u-r\right) }=\prod \left( u-i\right) ,i\neq s,i\neq r,s\neq r
\end{equation*}%
and $g_{s}\left( u\right) =\left( u-r\right) Q_{sr}\left( u\right) =\frac{%
f\left( u\right) }{\left( u-s\right) }$. By using these notations, the ring $%
R_{s,p}$ can be write under the form: 
\begin{equation*}
R_{s,p}=\mathbb{Z}_{p}[u]/\left( f\left( u\right) /\left( u-s\right) \right)
=\mathbb{Z}_{p}[u]/\left( g_{s}\left( u\right) \right) .
\end{equation*}%
We remark that $R_{s,p}=\underset{i\in I_{s}}{\prod }\mathbb{Z}%
_{p}[u]/\left( u-i\right) =\underset{p-1\text{ times}}{\underbrace{\mathbb{Z}%
_{p}\times ...\times \mathbb{Z}_{p}}}.$

For $r\in I_{s}$, we consider the element $q_{r}=Q_{sr}\left( r\right) ^{-1}$
\textit{mod }$p$.\medskip 

\textbf{Proposition 2.} \textit{By using the above notations, the following
statements are true:}

\textit{i) The element} $\alpha _{r}\left( u\right) =q_{r}Q_{sr}\left(
u\right) \,$\ \textit{is an idempotent in the ring} $R_{s,p}$.\medskip

\textit{ii)} $\alpha _{i}\left( u\right) \alpha _{j}\left( u\right) =0$%
\textit{, for} $i\neq j$ \textit{and} $\underset{j\in I_{s}}{\sum }\alpha
_{j}\left( u\right) =1.$

\textbf{Proof.} i) Indeed, we will prove that $\alpha _{r}^{2}\left(
u\right) =\alpha _{r}\left( u\right) $ \textit{mod} $g_{s}\left( u\right) ,$
for all $r\in I_{s}$. We have 
\begin{equation*}
\alpha _{r}^{2}\left( u\right) -\alpha _{r}\left( u\right)
=q_{r}Q_{sr}\left( u\right) ^{2}-q_{r}Q_{sr}\left( u\right) =
\end{equation*}%
\begin{equation*}
=q_{r}^{2}Q_{sr}^{2}\left( u\right) -q_{r}Q_{sr}\left( u\right)
=q_{r}Q_{sr}\left( u\right) \left( q_{r}Q_{sr}\left( u\right) -1\right) ,
\end{equation*}
then 
\begin{equation*}
Q_{sr}\left( u\right) \mid (\alpha _{r}^{2}\left( u\right) -\alpha
_{r}\left( u\right) ).
\end{equation*}
Now, we compute the elemet $\alpha _{r}^{2}\left( u\right) -\alpha
_{r}\left( u\right) $.

We obtain 
\begin{equation*}
\alpha _{r}^{2}\left( r\right) -\alpha _{r}\left( r\right)
=q_{r}Q_{sr}\left( r\right) \left( q_{r}Q_{sr}\left( r\right) -1\right) =
\end{equation*}%
\begin{equation*}
=Q_{sr}\left( j\right) ^{-1}Q_{sr}\left( r\right) (Q_{sr}\left( r\right)
^{-1}Q_{sr}\left( r\right) -1)=0.
\end{equation*}%
Then $\left( u-r\right) \mid (\alpha _{r}^{2}\left( u\right) -\alpha
_{r}\left( u\right) )$ and $g_{s}\left( u\right) \mid $ $(\alpha
_{r}^{2}\left( u\right) -\alpha _{r}\left( u\right) )$. We get $\alpha
_{r}^{2}\left( u\right) =\alpha _{r}\left( u\right) $ \textit{mod} $%
g_{s}\left( u\right) $ and it results that $\alpha _{r}\left( u\right) $ is
an nontrivial idempotent in $R_{s,p}$.

ii) It is clear that $\alpha _{i}\left( u\right) \alpha _{j}\left( u\right)
=0$. We must prove that $\underset{j\in I_{s}}{\sum }\alpha _{j}\left(
u\right) =1$. If $l\in I_{s}$, we have $\underset{j\in I_{s}}{\sum }\alpha
_{j}\left( l\right) =1$, since $\alpha _{l}\left( l\right) =Q_{sl}\left(
l\right) ^{-1}Q_{sl}\left( l\right) =1,$ if $j=l$ and $\alpha _{j}\left(
l\right) =0$, for $j\neq l$. Therefore, $\{\alpha _{j}\left( u\right)
\}_{j\in I_{s}}$ represents a set of orthogonal idempotents. $\Box \medskip $

From the above proposition, it results that we can apply the Peirce
decomposition, and the ring $R_{s,p}$ has the following form: 
\begin{equation}
R_{s,p}=\underset{i\in I_{s}}{\bigoplus }\alpha _{i}R_{s,p}.  \tag{1.}
\end{equation}

\textbf{Proposition 3.} \textit{Let} $\mathcal{C}$ \textit{be a linear code
of length} $n$ \textit{and dimension} $k$ \textit{over the ring} $R_{s.p}$.

\textit{i)} \textit{From relation }$\left( 1\right) $\textit{, we have} 
\begin{equation}
\mathcal{C}=\underset{i\in I_{s}}{\bigoplus }\alpha _{i}\mathcal{C}_{i}, 
\tag{2.}
\end{equation}%
\textit{where} $\mathcal{C}_{i},i\in I_{s},$ \textit{are linear codes of
length} $n$ \textit{and dimension} $k$ \textit{over the field} $\mathbb{Z}%
_{p}$.

\textit{ii)} $\left\vert \mathcal{C}\right\vert =\underset{i\in I_{s}}{\prod 
}\left\vert \mathcal{C}_{i}\right\vert =p^{\left( p-1\right) k}$.

\textit{iii)} \textit{If} $\mathcal{C}_{i}$ \textit{are linear code of length%
} $n$ \textit{and dimension} $k_{i}$ \textit{over the field} $\mathbb{Z}%
_{p},i\in I_{s},$ \textit{and} $G_{i}$ \textit{its generated matrix, we
obtain that the generated matrix of the code} $\mathcal{C}$ \textit{is of
the form}%
\begin{equation*}
G=\underset{i\in I_{s}}{\sum }\alpha _{i}G_{i}.
\end{equation*}

\textit{iv) For the linear code} $\mathcal{C}$\textit{, the minimum Hamming
distance, }$d_{H}\left( \mathcal{C}\right) $\textit{, has the following
formula:}%
\begin{equation*}
d_{H}\left( \mathcal{C}\right) =\underset{i\in I_{s}}{~\text{\textit{min}}}%
\{d_{H}\left( \mathcal{C}_{i}\right) \}
\end{equation*}

\bigskip

\textbf{Proof.} i) The element $x\in R_{s,p}$ has the following form $x=%
\underset{i=0}{\overset{p-1}{\sum }}x_{i}u^{i},x_{i}\in \mathbb{Z}_{p}$.
From here, by using the fact that $\{\alpha _{i}\}_{i\in I_{s}}$ are
orthogonal idempotents, we obtain $\alpha _{i}x=x_{i}\alpha _{i},x_{i}\in 
\mathbb{Z}_{p}$, for all $i\in I_{s}$. From here, it results that relation $%
\left( 2\right) $ is true.

ii) We have that $\left\vert \mathcal{C}_{i}\right\vert =p^{k}$, therefore $%
\underset{i\in I_{s}}{\prod }\left\vert \mathcal{C}_{i}\right\vert
=p^{\left( p-1\right) k}$. Since $\mathcal{C}$ is a linear code of length $n$
\textit{a}nd dimension $k$ over the ring $R_{s.p}$, we have $\left\vert 
\mathcal{C}\right\vert =\left\vert R_{s.p}\right\vert ^{k}=(p^{\left(
p-1\right) })^{k}=p^{\left( p-1\right) k}$.

iii) It is obvious from the above results.

iv) It is obvious from the definition of Hamming distance. $\Box \medskip $

\bigskip \textbf{Remark 4.} It is clear that the linear codes $\mathcal{C}%
_{i}$ are the form: $\mathcal{C}_{i}=\{x_{i}\in \mathbb{Z}_{p}^{n},$ such
that there are $x_{j}\in \mathbb{Z}_{p}^{n}$, for all $j\in I_{s}-\{i\}$,
with $\underset{i=0}{\overset{p-1}{\sum }}x_{i}u^{i}\in \mathcal{C}\},i\in
I_{s}$.\medskip

\textbf{Proposition 5. }\textit{Let} $\mathcal{C}=\underset{i\in I_{s}}{%
\bigoplus }\alpha _{i}\mathcal{C}_{i}$ \textit{be a linear code of length} $%
n $ \textit{over the ring} $R_{s,p}$. \textit{Therefore, its dual code has
the following formula:} 
\begin{equation*}
\mathcal{C}^{\perp }=\underset{i\in I_{s}}{\bigoplus }\alpha _{i}\mathcal{C}%
_{i}^{\perp }.
\end{equation*}%
\textbf{Proof.} Indeed, let $c\in \mathcal{C}$, $c=\underset{i\in I_{s}}{%
\bigoplus }\alpha _{i}c_{i}$. We denote $\mathcal{D}=\underset{i\in I_{s}}{%
\bigoplus }\alpha _{i}\mathcal{C}_{i}^{\perp }$. Let $d\in \underset{i\in
I_{s}}{\bigoplus }\alpha _{i}\mathcal{C}_{i}^{\perp },d=\underset{i\in I_{s}}%
{\bigoplus }\alpha _{i}d_{i}$. We compute $c\cdot d=\underset{i\in I_{s}}{%
\prod }\alpha _{i}c_{i}d_{i}=0$, therefore $c\cdot d=0$. From here, it
results that $\mathcal{D\subseteq C}^{\perp }$. Since the ring $R_{s,p}$ is
a Frobenius ring, then $\left\vert \mathcal{C}\right\vert \left\vert 
\mathcal{C}^{\perp }\right\vert =\left\vert R_{s,p}\right\vert ^{n}$. From
here, we have $\left\vert \mathcal{D}\right\vert =\underset{i\in I_{s}}{%
\prod }\left\vert \mathcal{C}_{i}^{\perp }\right\vert =\underset{i\in I_{s}}{%
\prod }\frac{\left\vert \mathbb{Z}_{p}^{n}\right\vert }{\left\vert \mathcal{C%
}_{i}\right\vert }=\underset{i\in I_{s}}{\prod }\frac{p^{n}}{p^{k_{i}}}=%
\frac{p^{n\left( p-1\right) }}{\underset{i\in I_{s}}{\prod }p^{k_{i}}}=\frac{%
\left\vert R_{s,p}\right\vert ^{n}}{\left\vert \mathcal{C}\right\vert }%
=\left\vert \mathcal{C}^{\perp }\right\vert $. Therefore, $\mathcal{D}=%
\mathcal{C}^{\perp }$.$\Box \medskip \medskip \medskip $

In the following, we consider the ring $R_{s,p,n}=R_{s,p}[x]/\left(
x^{n}-1\right) $.\medskip\ 

\textbf{Remark 6.} A linear code $\mathcal{C}$ of length $n$ over the ring $%
R_{s,p}$ is a cyclic code of length $n$ over the ring $R_{s,p}$ if and only
if $\mathcal{C}$ is an ideal in \ $R_{s,p,n}$, or, equivalently, if $%
c=\left( c_{0},c_{1},...,c_{n-1}\right) \in \mathcal{C}$, then $c=\left(
c_{n-1},c_{0},c_{1},...,c_{n-2}\right) \in \mathcal{C}$.\medskip\ 

\textbf{Proposition 7.} \textit{Let} $\mathcal{C}=\underset{i\in I_{s}}{%
\bigoplus }\alpha _{i}\mathcal{C}_{i}$ \textit{be a linear code over the ring%
} $R_{s,p}$. \textit{Then} $\mathcal{C}=\underset{i\in I_{s}}{\bigoplus }%
\alpha _{i}\mathcal{C}_{i}$ \textit{is a cyclic code of length} $n$ \textit{%
over the ring }$R_{s,p}$ \textit{if and only if the codes} $\mathcal{C}_{i}$ 
\textit{are cyclic codes of length} $n$ \textit{over the ring} $\mathbb{Z}%
_{p}[x]/\left( x^{n}-1\right) ,$ \textit{for all} $i\in I_{s}$.\medskip\ 

\textbf{Proof.} Indeed, we consider the codewords $c_{i}=\left(
c_{i,0},c_{i,1},...,c_{i,n-1}\right) \in \mathcal{C}_{i},i\in I_{s}$ and $%
c_{j}=\underset{i\in I_{s}}{\sum }\alpha _{i}c_{ij},j\in \{0,1,...,n-1\}$.
From here, we obtain that $c=\left( c_{0},c_{1},...,c_{n-1}\right) \in 
\mathcal{C}$. Since $\mathcal{C}$ is a cyclic code, we have that $\left(
c_{n-1},c_{0},c_{1},...,c_{n-2}\right) \in \mathcal{C}$. It results $\left(
c_{n-1},c_{0},c_{1},...,c_{n-2}\right) =\underset{i\in I_{s}}{\sum }\alpha
_{i}\left( c_{i,n-1},c_{i,0},...,c_{i,n-2}\right) $. From relation $\left(
2\right) $, we obtain the unicity of the linear codes decomposition over the
ring $R_{s,p}$, then $\left( c_{i,n-1},c_{i,0},...,c_{i,n-2}\right) \in $ $%
\mathcal{C}_{i}$ and $\mathcal{C}_{i}$ is a cyclic code for all $i\in I_{s}$.

For the converse statement, we consider that $\mathcal{C}_{i}$ are cyclic
codes over the field $\mathbb{Z}_{p}$, for all $i\in I_{s}$, and let $%
c=\left( c_{0},c_{1},...,c_{n-1}\right) \in \mathcal{C}$. Since

$c=\underset{i\in I_{s}}{\sum }\alpha _{i}\left(
c_{i,0},...,c_{i,n-2},c_{i,n-1}\right) \in $ $\underset{i\in I_{s}}{%
\bigoplus }\alpha _{i}\mathcal{C}_{i}$, it results \ that the linear code $c$
is a cyclic code of length $n$ over the ring $R_{s,p}.\Box \medskip $

\begin{equation*}
\end{equation*}

\textbf{3.} \textbf{Example}%
\begin{equation*}
\end{equation*}

\bigskip \textbf{- Linear codes}

Let $p=5$ and $f\left( u\right) =u^{5}-u=u\left( u-1\right) \left(
u-2\right) \left( u-3\right) \left( u-4\right) $. We choose $s=4$ and we
define $R_{4,5}=\mathbb{Z}_{5}[u]/\left( u\left( u-1\right) (u-2)\left(
u+2\right) \right) $. This ring has cardinal $5^{4}$. We denote with $%
I_{4}=\{0,1,2,3\}$. We have 
\begin{equation*}
Q_{4r}\left( u\right) =\frac{u\left( u-1\right) \left( u-2\right) \left(
u-3\right) \left( u-4\right) }{\left( u-4\right) \left( u-r\right) }=\prod
\left( u-i\right) ,i\neq 4,i\neq r,r\neq 4.
\end{equation*}%
We consider $g_{4}\left( u\right) =\left( u-r\right) Q_{4r}\left( u\right) =%
\frac{f\left( u\right) }{\left( u-4\right) }$. Then $R_{4,5}=\mathbb{Z}%
_{5}[u]/\left( u\left( u-1\right) \left( u-2\right) \left( u-3\right)
\right) $. We denote with $q_{r}=Q_{4r}\left( r\right) ^{-1}$ \textit{mod }$%
5 $, $r\in I_{4}$. The element $\alpha _{r}\left( u\right)
=q_{r}Q_{4r}\left( u\right) \,$\ is an idempotent in the ring $R_{4,5}$. We
have the following idempotents: $\alpha _{0}\left( u\right)
=q_{0}Q_{40}\left( u\right) =4\left( u-1\right) \left( u-2\right) \left(
u-3\right) $; $\alpha _{1}\left( u\right) =q_{1}Q_{41}\left( u\right)
=3u\left( u-2\right) \left( u-3\right) $; $\alpha _{2}\left( u\right)
=q_{2}Q_{42}\left( u\right) =2u\left( u-1\right) \left( u-3\right) $; $%
\alpha _{3}\left( u\right) =q_{3}Q_{43}\left( u\right) =u\left( u-1\right)
\left( u-2\right) $.

i) We consider a linear code of length $5$ and dimesion $3$ given by the
following generator matrix 
\begin{equation*}
G=\left( 
\begin{array}{ccccc}
1 & 0 & 0 & u^{3}+3u^{2}+2u+1 & u^{3}+u^{2}+4u+1 \\ 
0 & 1 & 0 & 3u^{3}+3u^{2}+4u+2 & 2u^{3}+u^{2}+u \\ 
0 & 0 & 1 & 4u^{3}+4u^{2}+2u & 4u^{3}+2u^{2}+u+3%
\end{array}%
\right) ,
\end{equation*}%
$\mathcal{C}=\alpha _{0}\mathcal{C}_{0}\oplus \alpha _{1}\mathcal{C}%
_{1}\oplus \alpha _{2}\mathcal{C}_{2}\oplus \alpha _{3}\mathcal{C}_{3}$.

We have that 
\begin{equation*}
G=\left( 
\begin{array}{ccccc}
1 & 0 & 0 & \alpha _{0}+2\alpha _{1}+\alpha _{3} & \alpha _{0}+2\alpha
_{1}+\alpha _{2}+4\alpha _{3} \\ 
0 & 1 & 0 & 2\alpha _{0}+2\alpha _{1}+\alpha _{2}+2\alpha _{3} & 4\alpha
_{1}+2\alpha _{2}+\alpha _{3} \\ 
0 & 0 & 1 & 2\alpha _{2} & 3\alpha _{0}+2\alpha _{3}%
\end{array}%
\right) .
\end{equation*}

Therefore, $G=\alpha _{0}\left( 
\begin{array}{ccccc}
1 & 0 & 0 & 1 & 1 \\ 
0 & 1 & 0 & 2 & 0 \\ 
0 & 0 & 1 & 0 & 3%
\end{array}%
\right) $+$\alpha _{1}\left( 
\begin{array}{ccccc}
1 & 0 & 0 & 2 & 2 \\ 
0 & 1 & 0 & 2 & 4 \\ 
0 & 0 & 1 & 0 & 0%
\end{array}%
\right) $+\newline
+$\alpha _{2}\left( 
\begin{array}{ccccc}
1 & 0 & 0 & 0 & 1 \\ 
0 & 1 & 0 & 1 & 2 \\ 
0 & 0 & 1 & 2 & 0%
\end{array}%
\right) $+$\alpha _{3}\left( 
\begin{array}{ccccc}
1 & 0 & 0 & 1 & 4 \\ 
0 & 1 & 0 & 2 & 1 \\ 
0 & 0 & 1 & 0 & 2%
\end{array}%
\right) $, where $G_{0}=\left( 
\begin{array}{ccccc}
1 & 0 & 0 & 1 & 1 \\ 
0 & 1 & 0 & 2 & 0 \\ 
0 & 0 & 1 & 0 & 3%
\end{array}%
\right) $ is the generator matrix for the linear code $\mathcal{C}_{0}$, $%
G_{1}=\left( 
\begin{array}{ccccc}
1 & 0 & 0 & 2 & 2 \\ 
0 & 1 & 0 & 2 & 4 \\ 
0 & 0 & 1 & 0 & 0%
\end{array}%
\right) $ s the generator matrix for the linear code $\mathcal{C}_{1}$, $%
G_{2}=\left( 
\begin{array}{ccccc}
1 & 0 & 0 & 0 & 1 \\ 
0 & 1 & 0 & 1 & 2 \\ 
0 & 0 & 1 & 2 & 0%
\end{array}%
\right) $ is the generator matrix for the linear code $\mathcal{C}_{2}$, $%
G_{3}=\left( 
\begin{array}{ccccc}
1 & 0 & 0 & 1 & 4 \\ 
0 & 1 & 0 & 2 & 1 \\ 
0 & 0 & 1 & 0 & 2%
\end{array}%
\right) $ is the generator matrix for the linear code $\mathcal{C}_{3}$.

We remark that $d_{H}\left( \mathcal{C}_{0}\right) =2$, $d_{H}\left( 
\mathcal{C}_{1}\right) =1$, $d_{H}\left( \mathcal{C}_{2}\right) =d_{H}\left( 
\mathcal{C}_{3}\right) =2$, therefore $d_{H}\left( \mathcal{C}\right) =1$.

ii) We consider the linear code 
\begin{equation*}
G=\left( 
\begin{array}{ccccc}
1 & 0 & 0 & 1 & 1 \\ 
0 & 1 & 0 & 1 & 2 \\ 
0 & 0 & 1 & 3u^{3}+4u^{2}+3u+1 & 4u^{3}+u^{2}+u+3%
\end{array}%
\right) ,
\end{equation*}%
$\mathcal{C}=\alpha _{0}\mathcal{C}_{0}\oplus \alpha _{1}\mathcal{C}%
_{1}\oplus \alpha _{2}\mathcal{C}_{2}\oplus \alpha _{3}\mathcal{C}_{3}$.

We have that 
\begin{equation*}
G=\left( 
\begin{array}{ccccc}
1 & 0 & 0 & \alpha _{0}+\alpha _{1}+\alpha _{2}+\alpha _{3} & \alpha
_{0}+\alpha _{1}+\alpha _{2}+\alpha _{3} \\ 
0 & 1 & 0 & \alpha _{0}+\alpha _{1}+\alpha _{2}+\alpha _{3} & 2\alpha
_{0}+2\alpha _{1}+2\alpha _{2}+2\alpha _{3} \\ 
0 & 0 & 1 & \alpha _{0}+\alpha _{1}+2\alpha _{2}+2\alpha _{3} & 3\alpha
_{0}+4\alpha _{1}+\alpha _{2}+3\alpha _{3}%
\end{array}%
\right) .
\end{equation*}%
Therefore, $G=\alpha _{0}\left( 
\begin{array}{ccccc}
1 & 0 & 0 & 1 & 1 \\ 
0 & 1 & 0 & 1 & 2 \\ 
0 & 0 & 1 & 1 & 3%
\end{array}%
\right) $+$\alpha _{1}\left( 
\begin{array}{ccccc}
1 & 0 & 0 & 1 & 1 \\ 
0 & 1 & 0 & 1 & 2 \\ 
0 & 0 & 1 & 1 & 4%
\end{array}%
\right) $+\newline
+$\alpha _{2}\left( 
\begin{array}{ccccc}
1 & 0 & 0 & 1 & 1 \\ 
0 & 1 & 0 & 1 & 2 \\ 
0 & 0 & 1 & 2 & 1%
\end{array}%
\right) $+$\alpha _{3}\left( 
\begin{array}{ccccc}
1 & 0 & 0 & 1 & 1 \\ 
0 & 1 & 0 & 1 & 2 \\ 
0 & 0 & 1 & 2 & 3%
\end{array}%
\right) $, where $G_{0}=\left( 
\begin{array}{ccccc}
1 & 0 & 0 & 1 & 1 \\ 
0 & 1 & 0 & 1 & 2 \\ 
0 & 0 & 1 & 1 & 3%
\end{array}%
\right) $ is the generator matrix for the linear code $\mathcal{C}_{0}$, $%
G_{1}=\left( 
\begin{array}{ccccc}
1 & 0 & 0 & 1 & 1 \\ 
0 & 1 & 0 & 1 & 2 \\ 
0 & 0 & 1 & 1 & 4%
\end{array}%
\right) $ s the generator matrix for the linear code $\mathcal{C}_{1}$, $%
G_{2}=\left( 
\begin{array}{ccccc}
1 & 0 & 0 & 1 & 1 \\ 
0 & 1 & 0 & 1 & 2 \\ 
0 & 0 & 1 & 2 & 1%
\end{array}%
\right) $ is the generator matrix for the linear code $\mathcal{C}_{2}$, $%
G_{3}=\left( 
\begin{array}{ccccc}
1 & 0 & 0 & 1 & 1 \\ 
0 & 1 & 0 & 1 & 2 \\ 
0 & 0 & 1 & 2 & 3%
\end{array}%
\right) $ is the generator matrix for the linear code $\mathcal{C}_{3}$.

We remark that $d_{H}\left( \mathcal{C}_{0}\right) =d_{H}\left( \mathcal{C}%
_{1}\right) =d_{H}\left( \mathcal{C}_{2}\right) =d_{H}\left( \mathcal{C}%
_{3}\right) =3$, therefore $d_{H}\left( \mathcal{C}\right) =3$.This code is
an MDS code (Maximum Distance Separable code).

\textbf{-Dual codes}

i) For the dual code, we know that if the generator matrix $G$ has the form $%
\left( I_{3}\shortmid M\right) $, then the parity check matrix $H$ is of the
form $\left( -M^{t}\shortmid I_{2}\right) $. Since $G_{0}=\left(
I_{3}\shortmid M_{0}\right) $, we obtain $H_{0}=\left( 
\begin{array}{ccccc}
4 & 3 & 0 & 1 & 0 \\ 
4 & 0 & 2 & 0 & 1%
\end{array}%
\right) $. From $G_{1}=\left( 
\begin{array}{ccccc}
1 & 0 & 0 & 2 & 2 \\ 
0 & 1 & 0 & 2 & 4 \\ 
0 & 0 & 1 & 0 & 0%
\end{array}%
\right) $, we get $H_{1}=\left( 
\begin{array}{ccccc}
3 & 3 & 0 & 1 & 0 \\ 
3 & 1 & 0 & 0 & 1%
\end{array}%
\right) $; to $G_{2}=\left( 
\begin{array}{ccccc}
1 & 0 & 0 & 0 & 1 \\ 
0 & 1 & 0 & 1 & 2 \\ 
0 & 0 & 1 & 2 & 0%
\end{array}%
\right) $ it corresponds $H_{2}=\left( 
\begin{array}{ccccc}
0 & 4 & 3 & 1 & 0 \\ 
4 & 3 & 0 & 0 & 1%
\end{array}%
\right) $ and for $G_{3}=\left( 
\begin{array}{ccccc}
1 & 0 & 0 & 1 & 4 \\ 
0 & 1 & 0 & 2 & 1 \\ 
0 & 0 & 1 & 0 & 2%
\end{array}%
\right) $ we obtain $H_{3}=\left( 
\begin{array}{ccccc}
4 & 3 & 0 & 1 & 0 \\ 
1 & 4 & 3 & 0 & 1%
\end{array}%
\right) $. Therefore, the matrix of the dual code $\mathcal{C}^{\perp }$ is 
\begin{equation*}
H=\left( 
\begin{array}{ccccc}
4\alpha _{0}+3\alpha _{1}+4\alpha _{3} & 3\alpha _{0}+3\alpha _{1}+4\alpha
_{2}+3\alpha _{3} & 3\alpha _{2} & 1 & 0 \\ 
4\alpha _{0}+3\alpha _{1}+4\alpha _{2}+\alpha _{3} & \alpha _{1}+3\alpha
_{2}+4\alpha _{3} & 2\alpha _{0}+3\alpha _{3} & 0 & 1%
\end{array}%
\right) =
\end{equation*}%
\begin{equation*}
=\left( 
\begin{array}{ccccc}
4u^{3}+2u^{2}+3u+4 & 2u^{3}+2u^{2}+u+3 & u^{3}+u^{2}+3u & 1 & 0 \\ 
4\alpha _{0}+3\alpha _{1}+4\alpha _{2}+\alpha _{3} & 3u^{3}+4u^{2}+4u & 
u^{3}+3u^{2}+4u+2 & 0 & 1%
\end{array}%
\right) \text{.}
\end{equation*}

ii) In this case,for the dual code, we obtain: for $G_{0}$, we have $%
H_{0}=\left( 
\begin{array}{ccccc}
4 & 4 & 4 & 1 & 0 \\ 
4 & 3 & 2 & 0 & 1%
\end{array}%
\right) $; for $G_{1}$, we obtain $H_{1}=\left( 
\begin{array}{ccccc}
4 & 4 & 4 & 1 & 0 \\ 
4 & 3 & 1 & 0 & 1%
\end{array}%
\right) $; for $G_{2}$, it results $H_{2}=\left( 
\begin{array}{ccccc}
4 & 4 & 3 & 1 & 0 \\ 
4 & 3 & 4 & 0 & 1%
\end{array}%
\right) $; for $G_{3}$, we get $H_{3}=\left( 
\begin{array}{ccccc}
4 & 4 & 3 & 1 & 0 \\ 
4 & 3 & 2 & 0 & 1%
\end{array}%
\right) $.

Therefore, the matrix of the dual code $\mathcal{C}^{\perp }$ is 
\begin{equation*}
H=\left( 
\begin{array}{ccccc}
4\alpha _{0}+4\alpha _{1}+4\alpha _{2}+4\alpha _{3} & 4\alpha _{0}+4\alpha
_{1}+4\alpha _{2}+4\alpha _{3} & 4\alpha _{0}+4\alpha _{1}+3\alpha
_{2}+3\alpha _{3} & 1 & 0 \\ 
4\alpha _{0}+4\alpha _{1}+4\alpha _{2}+4\alpha _{3} & 3\alpha _{0}+3\alpha
_{1}+3\alpha _{2}+3\alpha _{3} & 2\alpha _{0}+\alpha _{1}+4\alpha
_{2}+2\alpha _{3} & 0 & 1%
\end{array}%
\right) =
\end{equation*}%
\begin{equation*}
=\left( 
\begin{array}{ccccc}
4 & 4 & 2u^{3}+u^{2}+2u+4 & 1 & 0 \\ 
4 & 3 & u^{3}+4u^{2}+4u+2 & 0 & 1%
\end{array}%
\right) \text{.}
\end{equation*}

-\textbf{A cyclic code of length }$4$\textbf{\ over the ring }$R_{4,5}$. We
have that $x^{4}-1=\left( x-1\right) \left( x-2\right) \left( x-3\right)
\left( x-4\right) $. We consider $g_{0}\left( x\right) =x-2,g_{1}\left(
x\right) =x-3,g_{2}\left( x\right) =\left( x-1\right) \left( x-2\right)
=\allowbreak x^{2}-3x+2,g_{3}\left( x\right) =\left( x-4\right) $.
Therefore, the generator polynomial for the code $\mathcal{C}$ is 
\begin{equation*}
g\left( x\right) =\alpha _{0}g_{0}\left( x\right) +\alpha _{1}g_{1}\left(
x\right) +\alpha _{2}g_{2}\left( x\right) +\alpha _{3}g_{3}\left( x\right) =
\end{equation*}%
\begin{equation*}
=\left( 2u^{3}+2u^{2}+u\right) x^{2}+\left( 2u^{3}+2u^{2}+u+1\right)
x+\left( 3u^{3}+4u^{2}+2u+3\right) .
\end{equation*}

-\textbf{A cyclic code of length }$5$\textbf{\ over the ring }$R_{4,5}$. We
have $x^{5}-1=\left( x-1\right) ^{5}$ \textit{mod} $5$. We consider $%
g_{0}\left( x\right) =x-1=x+4$, $g_{1}\left( x\right)
=(x-1)^{3}=x^{3}+2x^{2}+3x+4$, $g_{2}\left( x\right) =\left( x-1\right)
^{2}=x^{2}+3x+1$, $g_{3}\left( x\right) =\left( x-1\right)
^{4}=x^{4}+x^{3}+x^{2}+x+1$. Therefore, the generator polynomial for the
code $\mathcal{C}$ is%
\begin{equation*}
g\left( x\right) \text{=}\left( u^{3}\text{+}2u^{2}\text{+}2u\right) x^{4}%
\text{+}\left( 4u^{3}\text{+}2u^{2}\right) x^{3}\text{+}\left( 4u^{3}\text{+}%
4u^{2}\text{+}4u\right) x^{2}\text{+}\left( 4u^{2}\text{+}3u\text{+}1\right)
x\text{+}\left( u^{3}\text{+}3u^{2}\text{+}u\text{+}4\right) .
\end{equation*}

\begin{equation*}
\end{equation*}

\textbf{Conclusions.} Studying linear and cyclic codes over finite rings has
led to efficient codes with optimal parameters. The algebraic properties of
finite rings make it easy to adapt algorithms to different noisy channels,
improving real-time error correction. By using ideals in polynomial rings,
cyclic codes keep hardware implementation very simple and fast. These
mathematical structures remain essential for secure data storage.

\begin{equation*}
\end{equation*}

\textbf{References}%
\begin{equation*}
\end{equation*}

\textbf{[AP; 05]} Andrade, A.A., Palazzo, R., \textit{Linear Codes over
Finite Rings}, TEMA Tend. Mat. Apl. Comput., 6(2) (2005), 207-217.

[\textbf{GNW; 04}] Greferath, M., Nechaev, A., Wisbauer, R., \textit{Finite
Quasi-Frobenius Modules and Linear Codes}, Journal of Algebra and Its
Applications, 3(3)(2004), 247-272.

[\textbf{H; 01}] Honold, T., \textit{Characterization of finite Frobenius
rings}, Arch. Math., 76 (2001), 406-415.

\textbf{[MGF; 18] }Ma, F., Gao, J., Fu, \ F.W.\textbf{,} \textit{%
Constacyclic codes over the ring} $\mathbb{F}_{q}+v\mathbb{F}_{q}+v^{2}%
\mathbb{F}_{q}$ \textit{and their applications of constructing new
non-binary quantum codes}, Quantum Inf Process (2018) 17:122,
https://doi.org/10.1007/s11128-018-1898-6.

\textbf{[MK; 04 ]} McCrimmon, K., \textit{A Taste of Jordan algebras},
Springer-Verlag New York, Inc., 2004, 562 p., ISBN 0-387-95447-3,
https://link.springer.com/chapter/10.1007/0-387-21796-7\_17.

[\textbf{SS; 16}] Sari, M., Siap, I., \textit{On quantum codes \ from cyclic
codes over a class of nonchain rings}, Bull. Korean Math. Soc. 53 (2016),
No. 6, pp. 1617--1628, http://dx.doi.org/10.4134/BKMS.b150544

\textbf{[W; 99]} Wood, J. A., \textit{Duality for modules over finite rings
and applications to coding theory}, American Journal of Mathematics,
121(1999), 555-575.

\textbf{[ZWS; 10]} Zhu, S. X. , Wang, Y., Shi, M. J., \textit{Cyclic codes
over} $\mathbb{F}_{2}$ $+v\mathbb{F}_{2}$, IEEE Trans. Inform. Theory,
56(4)(2010), 1680--1684.%
\begin{equation*}
\end{equation*}

Cristina Flaut$^{\ast }$(corresponding author)

{\small Faculty of Mathematics and Computer Science, Ovidius University,}

{\small Bd. Mamaia 124, 900527, Constan\c{t}a, Rom\^{a}nia,}

{\small \ http://www.univ-ovidius.ro/math/; https://www.cristinaflaut.com,}

{\small e-mail: cflaut@univ-ovidius.ro; cristina\_flaut@yahoo.com}

\bigskip

Bianca Liana Bercea-Straton

{\small PhD student at Doctoral School of Mathematics,}

{\small Ovidius University of Constan\c{t}a, Rom\^{a}nia}

{\small e-mail: biancaliana99@yahoo.com}

\end{document}